\documentclass[%
 reprint,
 amsmath,amssymb,
 aps,
]{revtex4-2}

\usepackage{graphicx}
\usepackage{dcolumn}
\usepackage{bm}
\usepackage[hidelinks]{hyperref}
\usepackage{array}

\begin{document}

\preprint{APS/123-QED}

\title{Frequency comb generation in active optomechanical cavities}

\author{Madhurendra Mishra}
\email{madhurendramishra24@gmail.com}
\affiliation{%
Department of Physics\\
Sri Guru Tegh Bahadur Khalsa College, University of Delhi\\
New Delhi -- 110007, India
}

\author{Adarsh Ganesan}
\email{adarsh@dubai.bits-pilani.ac.in}
\affiliation{%
Department of Electrical and Electronics Engineering\\
Birla Institute of Technology and Science Pilani, Dubai Campus\\
Dubai International Academic City, Dubai -- 345055, UAE
}

\begin{abstract}
We numerically investigate frequency-comb generation in an electrically pumped vertical-cavity surface-emitting laser coupled to two mechanical modes of a suspended mirror. Displacement-dependent cavity loss and radiation-pressure feedback couple the mechanical motion to the carrier and photon densities, producing harmonics and combination-frequency components in the photon-density spectrum. Mechanical quality factors primarily control the intensity of comb lines, whereas resonance frequencies determine their spectral distribution. The initial mechanical amplitudes influence the threshold current and relative spectral intensities, with the simultaneous excitation of both modes yielding a broader set of observable components. These results establish the dependence of the calculated comb structure on mechanical dissipation, resonance frequencies, and initial conditions in an active optomechanical cavity.
\end{abstract}

\maketitle


\section{Introduction}
Cavity optomechanics studies how a confined optical field interacts with mechanical motion, typically arising from the dependence of the optical resonance frequency on mechanical displacement~\cite{Aspelmeyer2014}. This interaction is inherently bidirectional: the optical field modifies the mechanical dynamics, and the mechanical motion, in turn, changes the optical response~\cite{Cohadon1999,Metzger2004}. This optomechanical interaction underlies a range of phenomena including regenerative oscillation~\cite{Kippenberg2005,Carmon2005}, the dynamical backaction and radiation-pressure cooling~\cite{Gigan2006,Arcizet2006,Schliesser2006,Chan2011}, the optomechanically induced transparency~\cite{SafaviNaeini2011}, and the coherent photon--phonon interactions~\cite{Verhagen2012}. Different cavity architectures have pushed these effects across a wide range of scales. Membrane-in-the-middle systems give strong dispersive coupling between optical and mechanical modes~\cite{Thompson2008}, and optomechanical crystals confine photons and phonons within the same nanostructure~\cite{Eichenfield2009}. Together, these platforms show that optomechanical resonators are nonlinear systems, where optical and mechanical modes exchange energy and evolve together.

These works have typically been conducted using passive cavities driven by a laser external to the device. A qualitatively different regime appears by placing the optical gain inside the cavity itself, such that the light is generated by the device rather than being fed into it. Yang et al. showed this directly in an electrically pumped VCSEL with a mechanically compliant mirror, where the optomechanical coupling drove regenerative oscillation and a periodic shift of the emitted wavelength~\cite{Yang2015}. Active-cavity backaction has since been used to demonstrate optomechanical self-cooling~\cite{Foley2018}, and mechanical motion inside an active cavity has been shown, more generally, to alter the lasing state itself: shifting its frequency, destabilizing it, or broadening its spectrum~\cite{Yu2022}.

Frequency-comb formation is one clear spectral signature of nonlinear oscillatory dynamics. An optical frequency comb -- a set of discrete evenly spaced spectral lines with fixed frequency relations -- has become a standard tool in precision metrology, spectroscopy, and frequency synthesis~\cite{Udem2002,Cundiff2003,Diddams2010}. Beyond mode-locked lasers, such combs can also form through nonlinear interactions in high-$Q$ microresonators~\cite{DelHaye2007,Kippenberg2011}, including dissipative-soliton states that yield coherent, low-noise spectra~\cite{Herr2014,Gaeta2019}. Mechanical resonators can develop the same kind of spectral structure. Phononic frequency combs were first predicted from nonlinear resonances~\cite{Cao2014} and later experimentally observed through intrinsic three-wave mixing in micromechanical resonators~\cite{Ganesan2017}. Later work traced the formation of the comb to four-wave mixing~\cite{Ganesan2017FWM}, multimode parametric coupling~\cite{Ganesan2018Parametric,Ganesan2018Coupled}, bifurcation dynamics~\cite{Czaplewski2018}, and coherent energy transfer between mechanical modes~\cite{Sun2023}. Internal resonance offers another route: combs have been reported in coupled micromechanical resonators near a resonance $1{:}3$~\cite{Wang2022Internal}, in MEMS resonators with internal resonance $1{:}2$~\cite{Gobat2023}, and in atomically thin NEMS resonators through modal interactions $1{:}1$ and $2{:}1$~\cite{Yousuf2023}. Comb formation can also persist under strong mechanical damping, including in fluid environments, when electromechanical coupling is present~\cite{Surappa2023}. Analytical work has mapped out the conditions for the formation of phononic combs~\cite{Qi2020}, and experiments have demonstrated single-mode nanomechanical combs~\cite{Ochs2022}, resonance tracking with phononic combs~\cite{Ganesan2019}, and mechanical overtone combs~\cite{deJong2023}. 

Optomechanical frequency combs arise from nonlinear interactions between optical fields and mechanical oscillations, producing spectral sidebands at harmonics and combination frequencies. Miri et al. analytically identified a parametric instability leading to comb formation in an optomechanical cavity~\cite{Miri2018}. Mercad\'e et al. experimentally demonstrated frequency comb generation through self-sustained mechanical oscillations in a silicon optomechanical crystal~\cite{Mercade2020}, while Hu et al. generated optical and microwave frequency combs through large-amplitude optomechanical oscillations~\cite{Hu2021}. Multimode coupling introduces additional spectral structure. Ng et al. demonstrated intermodulation between optical frequency combs generated by two simultaneously oscillating mechanical modes~\cite{Ng2023}, and Wang et al. investigated comb formation involving coupled mechanical modes and nonlinear dynamical regimes~\cite{Wang2024}. More recently, Wan et al. demonstrated comb generation through coupled thermal and optical forces in a nanoelectromechanical cavity~\cite{Wan2025}. Gou et al. reported a chip-scale silicon carbide optomechanical comb spanning 1--70 GHz with 42 phase-locked harmonics~\cite{Gou2025}, while Zhang et al. demonstrated mode-locked comb generation in a graphene--silica microresonator~\cite{Zhang2025}. These studies establish several mechanisms for the generation of combs in externally driven optomechanical cavities. 

While optomechanical frequency combs have been explored in passive cavities, in this work, we show that the laser output of a VCSEL--MEMS active optomechanical cavity settles into self-sustained oscillations that generate an optical frequency comb. Because the optical field is generated inside the same cavity as the mechanical motion, that motion acts directly on the laser dynamics rather than modulating a field supplied from outside. As a result, the emitted spectrum evolves into a frequency comb.

\section{Theory}
\label{sec:theory}

\subsection{Structure of a VCSEL--MEMS Device}

\begin{figure*}
    \centering
    \includegraphics[width=0.8\linewidth]{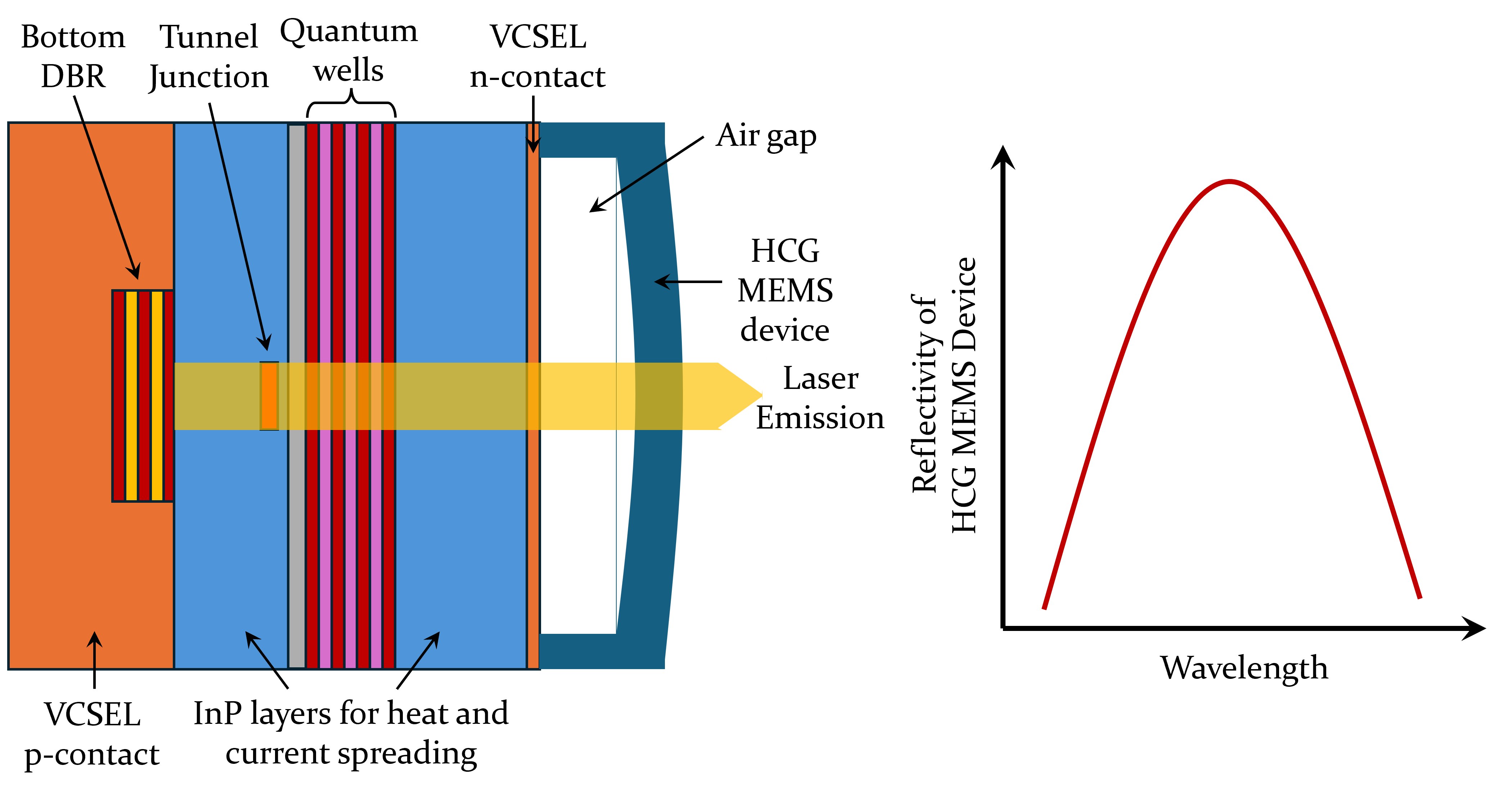}
    \caption{Schematic of the electrically pumped VCSEL--MEMS active optomechanical cavity.}
    \label{fig1}
\end{figure*}

Fig.~\ref{fig1} illustrates the structure and operating principle of the VCSEL--MEMS device. The optical cavity is bounded by a highly reflective distributed Bragg reflector (DBR) at one end and a suspended high-contrast grating (HCG) MEMS mirror at the other. Electrical injection supplies carriers to the active region of the quantum-well, where stimulated emission provides optical gain. The cavity mirrors confine the optical field through successive reflections, allowing amplification within the resonator while transmitting a fraction of the light as the laser output. The suspended high-contrast grating (HCG) mirror forms a mechanically compliant cavity boundary separated from the underlying structure by an air gap. Its oscillatory motion modulates the effective cavity length and shifts the resonant wavelength, thereby changing the wavelength-dependent HCG reflectivity and the associated photon loss rate $\kappa(x_{\mathrm{eff}})$. These changes modify the intracavity photon density $S(t)$ and, consequently, the radiation-pressure force acting on the suspended mirror. The resulting mechanical displacement feeds back into the optical response through the cavity wavelength and reflectivity, establishing a coupled dynamical system in which mechanical motion, photon density, and carrier dynamics evolve together.

\subsection{Coupled active-cavity model}

We model the VCSEL--MEMS device as an electrically pumped semiconductor laser whose suspended high-contrast grating (HCG) mirror supports two mechanical modes. Mirror motion changes the lasing wavelength and the HCG reflectivity, thereby changing the intracavity photon density and the radiation-pressure force. The two modes are coupled through this common optical variable, following the active-cavity framework~\cite{Yang2015}.

Let $x_j(t)$ be the displacement of mode $j=1,2$ from the static operating point. The effective displacement entering the optical response is
\begin{equation}
 x_{\rm eff}=\chi_1x_1+\chi_2x_2,\qquad \lambda(x_{\rm eff})=\lambda_b+a_\lambda x_{\rm eff},
 \label{eq:theory_wavelength}
\end{equation}
where $\chi_j$ are modal overlap factors, $\lambda_b$ is the bias wavelength, and $a_\lambda=d\lambda/dx_{\rm eff}$. In the local HCG range,
\begin{equation}
 R_1(\lambda)=R_{1p}+r_1(\lambda-\lambda_p)^2,
 \label{eq:theory_reflectivity}
\end{equation}
with peak reflectivity $R_{1p}$ at $\lambda_p$. The photon loss rate is
\begin{equation}
 \kappa(x_{\rm eff})=v_g\left[\alpha_i+\frac{1}{2L}\ln\!\left(\frac{1}{R_1[\lambda(x_{\rm eff})]R_2}\right)\right],
 \label{eq:theory_loss}
\end{equation}
where $v_g$ is the group velocity, $\alpha_i$ the internal loss, $L$ the effective cavity length, and $R_2$ the bottom-mirror reflectivity.

The carrier and photon densities, $N$ and $S$, obey
\begin{align}
 \dot N&=\frac{\eta I}{qV_a}-AN-BN^2-CN^3-\mathcal{G}(N,S),
 \label{eq:theory_carrier}\\
 \dot S&=\Gamma\mathcal{G}(N,S)-\kappa(x_{\rm eff})S+\Gamma\beta BN^2,
 \label{eq:theory_photon}\\
 \mathcal{G}(N,S)&=v_gg_0\frac{S(N-N_{\rm tr})}{1+\varepsilon S}.
 \label{eq:theory_gain}
\end{align}
Here $I$ is the injection current, $\eta$ the injection efficiency, $V_a$ the active volume, $g_0$ the differential gain, $N_{\rm tr}$ the transparency density, $\varepsilon$ the gain-compression coefficient, $\Gamma$ the confinement factor, and $\beta$ the spontaneous-emission coupling factor.

With $n_{\rm ph}=SV_a/\Gamma$, the radiation-pressure force is~\cite{Yang2015}
\begin{equation}
 F_{\rm RP}(S,x_{\rm eff})=\frac{hV_av_g}{\Gamma L}\frac{S}{\lambda(x_{\rm eff})}.
 \label{eq:theory_force}
\end{equation}
The two mechanical modes satisfy
\begin{equation}
\begin{aligned}
m_j\ddot{x}_j + b_j\dot{x}_j + m_j\omega_j^2 x_j
&= \xi_j\left[F_{\rm RP}(S,x_{\rm eff})-F_0\right],\\
b_j &= \frac{m_j\omega_j}{Q_j}.
\end{aligned}
\label{eq:theory_mechanics}
\end{equation}
where $\omega_j=2\pi f_j$, $m_j$ is the modal mass, $Q_j$ the mechanical quality factor, and $\xi_j$ the generalized-force factor. $F_0=F_{\rm RP}(S_0,0)$ is subtracted because the coordinates are measured from the statically displaced equilibrium. The modes interact through the common photon density; direct elastic and photothermal coupling are omitted.

\begin{figure*}
    \centering
    \includegraphics[width=\linewidth]{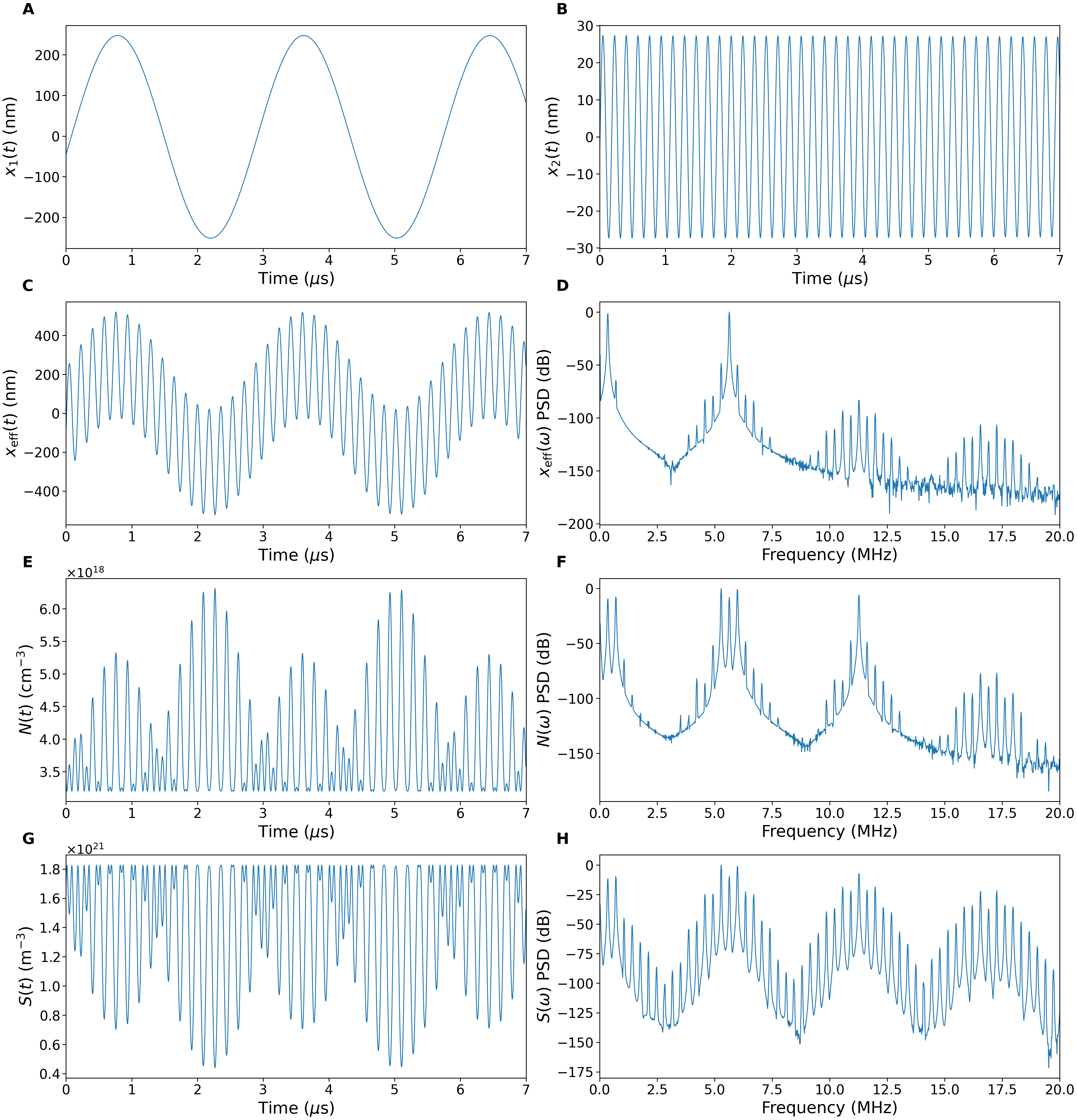}
    \caption{Coupled mechanical and laser dynamics at $f_1=353.6\,\mathrm{kHz}$, $f_2=5.636\,\mathrm{MHz}$, $Q_1=6000$, and $Q_2=10000$. (A) Displacement $x_1(t)$, (B) displacement $x_2(t)$, (C) effective displacement $x_{\mathrm{eff}}(t)=\chi_1x_1+\chi_2x_2$, and (D) its spectrum. (E) Carrier density $N(t)$ and (F) its spectrum. (G) Photon density $S(t)$ and (H) its spectrum, exhibiting frequency-comb structure.}
    \label{fig2}
\end{figure*}

\subsection{Optical-power waveform and frequency spectrum}

The simulated optical output is represented by $S(t)$. The final $7\,\mu\mathrm{s}$ of each trajectory is mapped to arbitrary units,
\begin{equation}
 P_{\rm AU}(t)=15\,\frac{S(t)-S_{\min}}{S_{\max}-S_{\min}},
 \label{eq:theory_power}
\end{equation}
which gives the normalized optical-power waveform used in the figures.

For the frequency-domain calculation, the final $50\,\mu\mathrm{s}$ is mean subtracted and multiplied by a Hann window $w_n$. For uniformly sampled values $S_n$ at sampling rate $f_s$, the one-sided periodogram is
\begin{equation}
\begin{aligned}
\mathcal{P}_S(f_k)
&=
\frac{2}{f_s\sum_n w_n^2}
\left|
\sum_{n=0}^{M-1}
w_n(S_n-\overline{S})
e^{-2\pi i kn/M}
\right|^2,\\
f_k&=\frac{kf_s}{M}.
\end{aligned}
\label{eq:theory_psd}
\end{equation}
The plotted spectrum is expressed in relative decibels, so the display shift changes amplitudes but not line positions.

The quadratic reflectivity produces nonlinear mixing. Substitution of Eq.~\eqref{eq:theory_wavelength} into Eq.~\eqref{eq:theory_reflectivity} gives
\begin{equation}
 R_1=R_{1b}+2r_1(\lambda_b-\lambda_p)a_\lambda x_{\rm eff}+r_1a_\lambda^2x_{\rm eff}^2,
 \label{eq:theory_mixing}
\end{equation}
and $x_{\rm eff}^2$ contains the product $2\chi_1\chi_2x_1x_2$. Thus the spectrum may contain harmonics, sum and difference tones, and higher-order components near
\begin{equation}
 f_{mn}=\left|m\widetilde f_1+n\widetilde f_2\right|,\qquad m,n\in\mathbb{Z},
 \label{eq:theory_combination}
\end{equation}
where $\widetilde f_j$ include any backaction-induced frequency shift. The calculated spectrum is the spectrum of the simulated intracavity intensity variable. A phase-resolved optical-field calculation would be required to establish field coherence~\cite{Hu2021}.

\subsection{Frequency and quality-factor sweeps}

\begin{figure*}
    \centering
    \includegraphics[width=\linewidth]{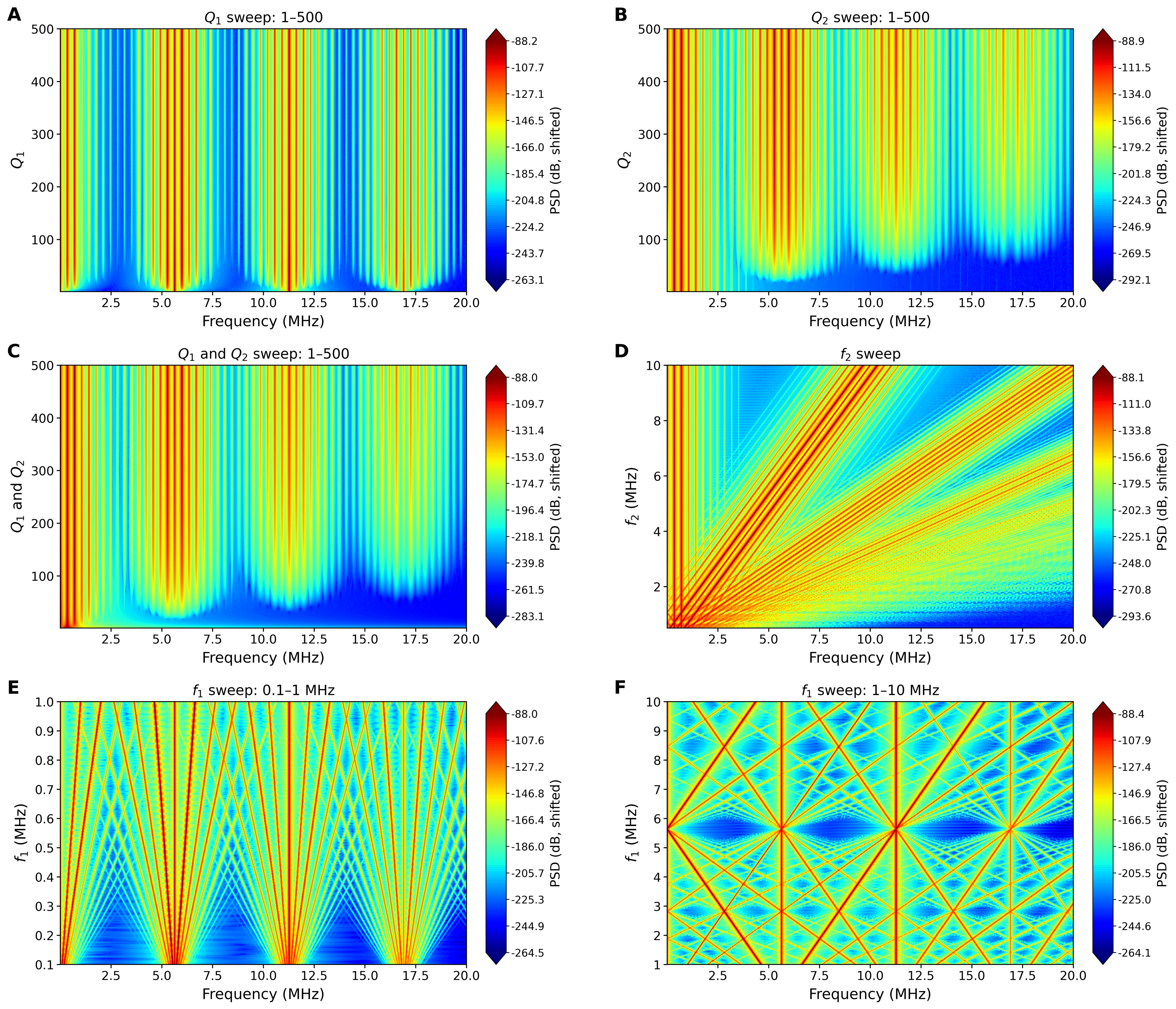}
    \caption{Dependence of the photon-density frequency comb on mechanical parameters. (A) $Q_1$ sweep, (B) $Q_2$ sweep, (C) simultaneous $Q_1=Q_2$ sweep, (D) $f_2$ sweep, (E) $f_1=0.1$--$1\,\mathrm{MHz}$ sweep, and (F) $f_1=1$--$10\,\mathrm{MHz}$ sweep.}
    \label{fig3}
\end{figure*}

\begin{table*}[htbp]
\caption{Fixed parameters used in the simulations.
The unit $1$ denotes a dimensionless quantity.}
\label{tab:theory_fixed}
\centering
\renewcommand{\arraystretch}{1.2}
\begin{tabular}{|c|l|c|c|}
\hline
\textbf{Symbol} & \textbf{Description} &
\textbf{Value} & \textbf{Unit} \\
\hline
$I$ & Injection current
& $18$ & $\mathrm{mA}$ \\
\hline
$\eta$ & Current injection efficiency
& $0.93$ & $1$ \\
\hline
$V_a$ & Active-region volume
& $12\times10^{-18}$ & $\mathrm{m^3}$ \\
\hline
$A$ & Nonradiative recombination coefficient
& $5.0\times10^7$ & $\mathrm{s^{-1}}$ \\
\hline
$B$ & Radiative recombination coefficient
& $8.0\times10^{-17}$ & $\mathrm{m^3\,s^{-1}}$ \\
\hline
$C$ & Auger recombination coefficient
& $3.5\times10^{-42}$ & $\mathrm{m^6\,s^{-1}}$ \\
\hline
$N_{\mathrm{tr}}$ & Transparency carrier density
& $8.0\times10^{23}$ & $\mathrm{m^{-3}}$ \\
\hline
$g_0$ & Differential gain coefficient
& $2.0\times10^{-20}$ & $\mathrm{m^2}$ \\
\hline
$\varepsilon$ & Gain-compression coefficient
& $2.0\times10^{-23}$ & $\mathrm{m^3}$ \\
\hline
$\Gamma$ & Optical confinement factor
& $0.025$ & $1$ \\
\hline
$\beta$ & Spontaneous-emission coupling factor
& $10^{-4}$ & $1$ \\
\hline
$\alpha_i$ & Internal optical loss coefficient
& $800$ & $\mathrm{m^{-1}}$ \\
\hline
$v_g$ & Optical group velocity
& $9.0\times10^7$ & $\mathrm{m\,s^{-1}}$ \\
\hline
$L$ & Effective cavity length
& $3.5$ & $\mu\mathrm{m}$ \\
\hline
$R_{1p}$ & Peak HCG reflectivity
& $0.9985$ & $1$ \\
\hline
$R_2$ & Bottom DBR-mirror reflectivity
& $0.999$ & $1$ \\
\hline
$r_1$ & Quadratic coefficient of HCG reflectivity
& $-2.0\times10^{13}$ & $\mathrm{m^{-2}}$ \\
\hline
$\lambda_p$ & Wavelength at peak HCG reflectivity
& $1560$ & $\mathrm{nm}$ \\
\hline
$\lambda_b$ & Bias lasing wavelength
& $1558$ & $\mathrm{nm}$ \\
\hline
$a_\lambda$ & Wavelength sensitivity to effective displacement
& $0.041$ & $1$ \\
\hline
$m_1$ & Effective mass of mechanical mode 1
& $200$ & $\mathrm{pg}$ \\
\hline
$m_2$ & Effective mass of mechanical mode 2
& $120$ & $\mathrm{pg}$ \\
\hline
$\chi_1$ & Displacement weight of mechanical mode 1
& $1$ & $1$ \\
\hline
$\chi_2$ & Displacement weight of mechanical mode 2
& $10$ & $1$ \\
\hline
$\xi_1$ & Radiation-pressure force weight of mode 1
& $1$ & $1$ \\
\hline
$\xi_2$ & Radiation-pressure force weight of mode 2
& $1$ & $1$ \\
\hline
\end{tabular}
\end{table*}

The reference point is $f_1=353.6\,\mathrm{kHz}$, $f_2=5.636\,\mathrm{MHz}$, $Q_1=6000$, and $Q_2=10000$. A frequency sweep changes $f_j$ and hence $\omega_j=2\pi f_j$, the mechanical stiffness $m_j\omega_j^2$, and, at fixed $Q_j$, the damping coefficient $b_j=m_j\omega_j/Q_j$. A quality-factor sweep changes the mechanical dissipation while keeping the corresponding resonance frequency fixed. The common-$Q$ scans impose $Q_1=Q_2=Q$.

All other parameters remain fixed. Each trajectory starts from the same laser steady state and prescribed mechanical initial conditions,
\begin{equation}
\begin{aligned}
x_j(0) &= x_{j,\rm amp}\cos\phi_j,\\
\dot{x}_j(0) &= -x_{j,\rm amp}\omega_j\sin\phi_j,
\end{aligned}
\label{eq:theory_initial}
\end{equation}
with $(x_{1,\rm amp},x_{2,\rm amp})=(250,30)\,\mathrm{nm}$ and $(\phi_1,\phi_2)=(0,0.20)$ radians. The equations are integrated for $60\,\mu\mathrm{s}$ using an implicit backward-differentiation method with relative tolerance $2\times10^{-6}$, maximum internal step $2\,\mathrm{ns}$, and sampling rate $250\,\mathrm{MHz}$. The final $50\,\mu\mathrm{s}$ gives the $20\,\mathrm{kHz}$ Fourier bins. The $2\,\mathrm{kHz}$ increase in sweep 2 is a parameter increase and does not imply $2\,\mathrm{kHz}$ spectral resolution. The waveform and spectrum are normalized independently for each trajectory; they are used to compare the temporal structure and relative spectral content.

\section{Results and Discussion}
\label{sec:results}

\begin{figure*}
    \centering
    \includegraphics[width=\linewidth]{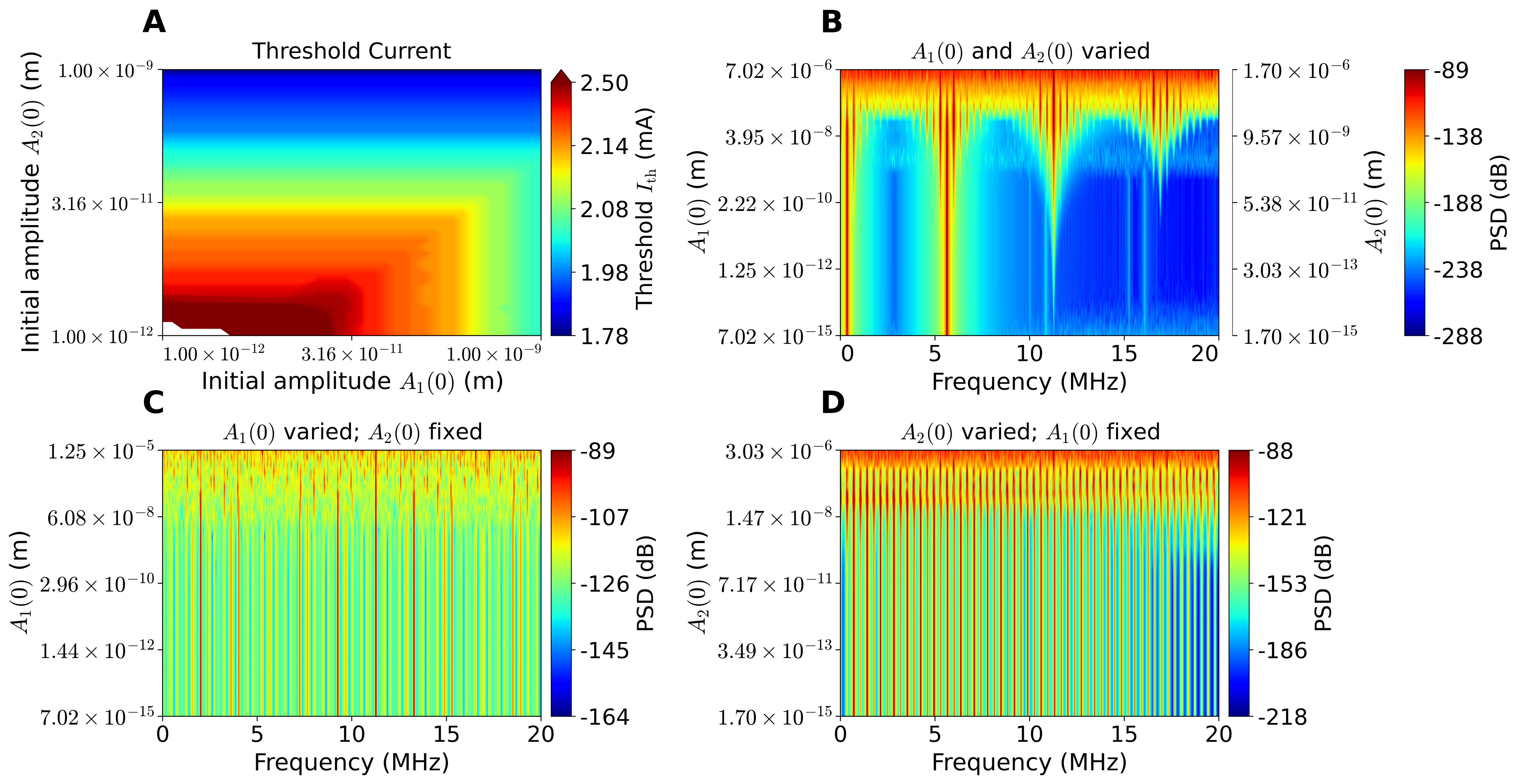}
    \caption{Dependence of the threshold current and photon-density frequency comb on the initial mechanical amplitudes. (A) Threshold current $I_{\mathrm{th}}$ as a function of $A_1(0)$ and $A_2(0)$ in a zoomed region of the parameter space. (B) Spectral evolution under simultaneous variation of $A_1(0)$ and $A_2(0)$. (C) Variation of $A_1(0)$ at fixed $A_2(0)$. (D) Variation of $A_2(0)$ at fixed $A_1(0)$.}
    \label{fig4}
\end{figure*}

At the reference operating point, $f_1=353.6\,\mathrm{kHz}$, $f_2=5.636\,\mathrm{MHz}$, $Q_1=6000$, and $Q_2=10000$, the two mechanical modes exhibit oscillations on distinct time scales, as shown in Figs.~\ref{fig2}(A) and \ref{fig2}(B). Their contributions combine through the effective displacement $x_{\mathrm{eff}}=\chi_1x_1+\chi_2x_2$, where the larger overlap factor $\chi_2$ makes the faster oscillation of the second mode prominent despite its smaller displacement amplitude, as illustrated in Fig.~\ref{fig2}(C). The resulting modulation of the lasing wavelength and HCG reflectivity changes the cavity loss and couples the mechanical motion to the carrier and photon densities, whose temporal responses are shown in Figs.~\ref{fig2}(E) and \ref{fig2}(G). The effective-displacement spectrum contains peaks near both mechanical frequencies together with additional spectral components, as shown in Fig.~\ref{fig2}(D). These features are consistent with harmonic mixing and combination-frequency arising from the nonlinear cavity response. The corresponding components appear with different relative intensities in the carrier and photon spectra in Figs.~\ref{fig2}(F) and \ref{fig2}(H). The photon-density spectrum exhibits a sequence of narrow lines extending across the calculated frequency range, with finer spacing associated with the lower-frequency mode and broader spectral groupings associated with the higher-frequency mode. This comb structure characterizes the simulated photon-density dynamics; establishing optical-field coherence requires a phase-resolved description.

The dependence on mechanical dissipation is examined through the quality-factor sweeps in Figs.~\ref{fig3}(A)--\ref{fig3}(C). Since $b_j=m_j\omega_j/Q_j$, increasing $Q_j$ reduces damping while keeping the prescribed mechanical resonance frequency fixed. The spectral lines consequently retain approximately the same positions, whereas their relative intensities vary. Increasing $Q_1$ primarily redistributes the strengths of existing components, with weaker lines becoming less discernible at lower quality factors, as shown in Fig.~\ref{fig3}(A). The $Q_2$ sweep produces a more pronounced spectral change. At low $Q_2$, the response is dominated by lower-frequency components, while many higher-frequency lines remain close to the spectral background. As $Q_2$ increases, these components become progressively more intense across the observation band, as shown in Fig.~\ref{fig3}(B). A similar recovery of higher-frequency components occurs when both quality factors are increased simultaneously in Fig.~\ref{fig3}(C). Thus, reducing mechanical dissipation generally improves the intensity of weaker comb lines, although the response differs between the two modes and is not uniform across frequencies.

The mechanical frequencies govern the positions and distribution of the spectral components, as illustrated in Figs.~\ref{fig3}(D)--\ref{fig3}(F). Varying $f_2$ shifts the spectral branches and changes their intersections within the observation window, producing a systematic reorganization of the comb structure in Fig.~\ref{fig3}(D). When $f_1$ is swept from $0.1$ to $1\,\mathrm{MHz}$, the separation and positions of the finer spectral features change, as shown in Fig.~\ref{fig3}(E). Extending the sweep to $1$--$10\,\mathrm{MHz}$ produces a more intricate pattern of sloped and intersecting branches in Fig.~\ref{fig3}(F), reflecting the changing relationships between the two mechanical frequencies. These features are consistent with the combination-frequency relation $f_{mn}=|m\widetilde f_1+n\widetilde f_2|$ in Eq.~\eqref{eq:theory_combination}. Unlike the quality-factor sweeps, frequency variation changes the locations of the spectral components and redistributes them across the observation band. Consequently, the number of comb lines within a fixed frequency interval depends on their positions and relative intensities rather than increasing monotonically with either mechanical frequency.

The influence of the initial mechanical conditions on the laser threshold and spectral response is examined in Fig.~\ref{fig4}. The threshold-current map in Fig.~\ref{fig4}(A) shows that $I_{\mathrm{th}}$ is highest when both initial displacement amplitudes, $A_1(0)$ and $A_2(0)$, are small. Increasing either amplitude generally reduces the calculated threshold, with the lowest values occurring toward larger $A_2(0)$. The threshold varies nonuniformly across the parameter space, indicating that its dependence cannot be described by either initial amplitude alone. The higher current required in the low-amplitude region reflects the threshold behavior of the coupled system under the prescribed initial conditions.

The corresponding spectral maps reveal distinct responses to simultaneous and independent variations of the initial amplitudes. When $A_1(0)$ and $A_2(0)$ increase together, additional spectral components become discernible, particularly at higher frequencies, while fewer lines remain observable in the low-amplitude region, as shown in Fig.~\ref{fig4}(B). Varying $A_1(0)$ alone while keeping $A_2(0)$ fixed mainly changes the relative intensities of the closely spaced spectral lines, without a clear monotonic increase in their number, as illustrated in Fig.~\ref{fig4}(C). In contrast, varying $A_2(0)$ at fixed $A_1(0)$ produces a stronger redistribution of the spectral intensity. Higher-frequency components are suppressed at smaller $A_2(0)$ and become more prominent as the amplitude increases, yielding a wider set of intense lines in Fig.~\ref{fig4}(D). This stronger dependence on the second mode is consistent with its larger displacement weight $\chi_2$ in $x_{\mathrm{eff}}$. The initial amplitudes therefore affect the intensity and relative strengths of the comb components, while the mechanical frequencies primarily determine their spectral positions. Because the spectral maps employ different color scales and display shifts, the comparison is restricted to the relative spectral structure within each panel rather than absolute power levels.

\section{Conclusion}
We have investigated frequency-comb generation in an electrically pumped VCSEL--MEMS active optomechanical cavity using a coupled model of two mechanical modes, carrier density $N(t)$, and photon density $S(t)$. The combined mechanical displacement modulates the cavity loss and produces a photon-density spectrum containing harmonics and combination-frequency components. The mechanical resonance frequencies $f_1$ and $f_2$ govern the spectral organization: the lower-frequency mode contributes to the finer comb-line spacing, while the higher-frequency mode shapes the broader spectral structure. Varying the mechanical quality factors $Q_1$ and $Q_2$ primarily changes the relative spectral intensities without substantially shifting the line positions. Stronger mechanical damping suppresses weaker spectral components, whereas increasing the quality factors allows additional lines to remain observable. Frequency sweeps further reveal the redistribution of spectral components as the mechanical resonance frequencies and their combinations change.

The initial mechanical amplitudes $A_1(0)$ and $A_2(0)$ influence both the threshold current $I_{\mathrm{th}}$ and the spectral response. The calculated threshold is higher when both initial amplitudes are small and generally decreases as they increase. Simultaneously increasing $A_1(0)$ and $A_2(0)$ enhances the intensity of spectral components, whereas varying either amplitude independently produces distinct changes in their relative intensities. In particular, reducing $A_2(0)$ suppresses higher-frequency components more strongly than the corresponding variation observed when $A_1(0)$ is varied alone. These results characterize how mechanical frequencies, dissipation, and initial conditions determine the frequency-comb structure within the coupled active-cavity model. Future investigations could examine the stability and multistability of the oscillatory states and incorporate a phase-resolved optical-field description to establish the coherence properties of the emitted spectrum.

\bibliography{apssamp}

\end{document}